\documentclass[]{spie}  

\usepackage{amsmath,amsfonts,amssymb}
\usepackage{graphicx}
\usepackage[colorlinks=true, allcolors=blue]{hyperref}

\title{COVID-19 detection from scarce chest x-ray image data using few-shot deep learning approach}

\author[*]{Shruti Jadon M.S.}
\affil{SPIE member; IEEE member; Sunnyvale, CA 95134 }
\authorinfo{Shruti Jadon, shrutijadon@ieee.org}

\begin{document} 
\maketitle

\begin{abstract}
In the current COVID-19 pandemic situation, there is an urgent need to screen infected patients quickly and accurately. Using deep learning models trained on chest X-ray images can become an efficient method for screening COVID-19 patients in these situations. Deep learning approaches are already widely used in the medical community. However, they require a large amount of data to be accurate. The open-source community \cite{zhao2020COVID-CT-Dataset} collectively has made efforts to collect and annotate the data, but it is not enough to train an accurate deep learning model. Few-shot learning \cite{jadon2020overview} is a sub-field of machine learning that aims to learn the objective with less amount of data. In this work, we have experimented with well-known solutions for data scarcity in deep learning to detect COVID-19. These include data augmentation, transfer learning, and few-shot learning, and unsupervised learning. We have also proposed a custom few-shot learning approach to detect COVID-19 using siamese networks \cite{yuan2017one}. Our experimental results showcased that we can implement an efficient and accurate deep learning model for COVID-19 detection by adopting the few-shot learning approaches even with less amount of data. Using our proposed approach we were able to achieve 96.4\% accuracy an improvement from 83\% using baseline models. Our code is available on github: \href{https://github.com/shruti-jadon/Covid-19-Detection}{https://github.com/shruti-jadon/Covid-19-Detection}
\end{abstract}


\section{Introduction}

A new coronavirus designated Covid-19 \cite{mo2020clinical} was first identified in Wuhan, the capital of China's Hubei province. It has been reported that people started developing pneumonia \cite{zhang2020clinical} without a clear cause and for which existing vaccines or treatments were not effective. The virus has shown evidence of human-to-human transmission. As of 24 January 2021, approximately 99 million people have contracted the virus and ~2 million have lost their lives. As ripple affect, a lot of people have lost their livelihood and about 40\% \cite{crayne2020traumatic} of small businesses have closed down. Majority of the countries weren't prepared for such pandemic situation in their hospitality domain which led to situation of a lot of doctors' risking their lives and working on multiple cases. With the help of technology, Covid-19 detection through CT scans can be automated to reduce up to ~2 minutes per scan basis which generally take close to 10-15 minutes. A lot of recent research papers \cite{afshar2020covid, cohen2020covid, horry2020covid} have suggested to tackle this issue with the help of deep learning, but with less amount of data and biased data scenarios, its tough to make a good inference on their results. 

In this work, we have experimented with some deep learning based techniques for training a model in low-data regime. We have also proposed a custom metrics based few-shot learning approach using siamese networks\cite{jadon2021improving}. Our proposed architecture has proven to perform well on scarce data. To validate the effectiveness and compare the performance of models, we have performed extensive set of experiments and showcased results. The paper is organized as follows: Section 2 explains the classification modeling approaches we have experimented with. In Section 3, we discuss about the evaluation metrics used to assess the performance of models on a given dataset. Our experimental results are listed in Section 4 on several real-world data-sets. We then finally conclude the outcomes in section 5. The project code has been made available for validation and replication of these experiments and can be found at: \href{https://github.com/shruti-jadon/Covid-19-Detection}{https://github.com/shruti-jadon/Covid-19-Detection}

  \begin{figure} [ht]
  \begin{center}
  \begin{tabular}{c} 
  \includegraphics[height=7.5cm]{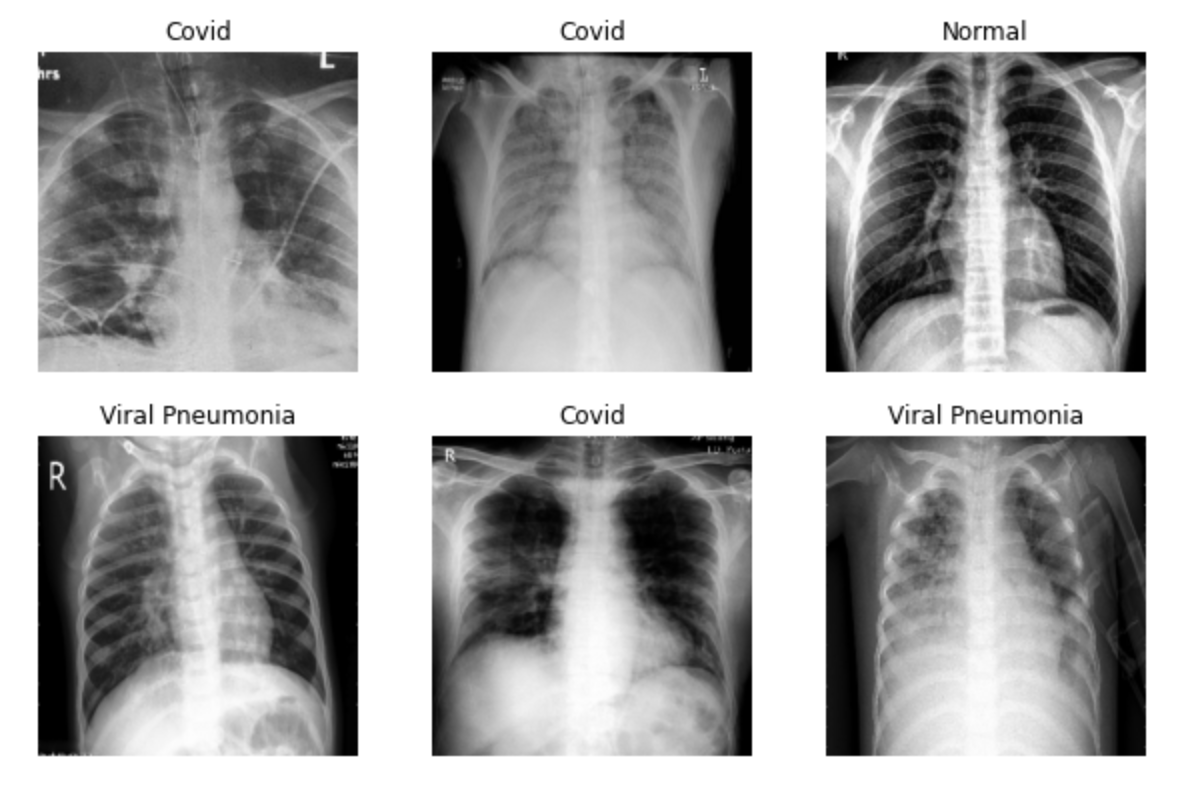}
	\end{tabular}
	\end{center}
  \caption[example] 
  { \label{fig:sample} 
Sample Covid-19 CT scan dataset \cite{cohen2020covid, zhao2020COVID-CT-Dataset} images} 
  \end{figure}

\subsection{Dataset}
For our research purposes, we have decided to experiment on two popular labeled datasets:
\begin{enumerate}
    \item \textbf{dataset-1:} Covid-19 Radiography database\cite{9144185} is a collaborative efforts by various universities in Asia. It consists of 1200 COVID-19 positive images, 1341 normal images, and 1345 viral pneumonia images. And,
    \item \textbf{dataset-2:} Covid-19 data collected with the help of University of Montreal \cite{cohen2020covid, zhao2020COVID-CT-Dataset}. This data consists of ~317 labeled images into three categories: Viral Pneumonia, Normal, and Covid.
\end{enumerate}
After analyzing the open-source data-set for Covid-19, we realized that it's a case of scarce data and therefore to train a high capacity model from scratch wouldn't be a good idea. To increase our dataset, we have taken help of data augmentation, but medical image augmentation have certain constraints, unlike generic vision based dataset we can't manipulate medical images. Therefore, we have used augmentation techniques such as shear, zoom, and rotation of smaller respective values.

\section{Modeling Approaches}
 The data scarcity situation is not new in the medical field. The medical community, in general, suffers from data-scarcity problems leading to the slow development of automation towards disease detection. The problem with less data is that if we train a good capacity model, we get underfitting, whereas if we train a low capacity model, our performance fails. To mitigate the effects of scarce data: the data augmentation approach generally works well, but if the data distribution is not a representation of real world data, it could lead to a biased model. For this research, we have selected certain widely used deep learning approaches to tackle scarce data situation.
\begin{enumerate}
\item Transfer Learning
\item Unsupervised Learning
\item Semi-Supervised Learning
\item Few-Shot Learning
\end{enumerate}

  \begin{figure} [!ht]
  \begin{center}
  \begin{tabular}{c} 
  \includegraphics[height=5cm]{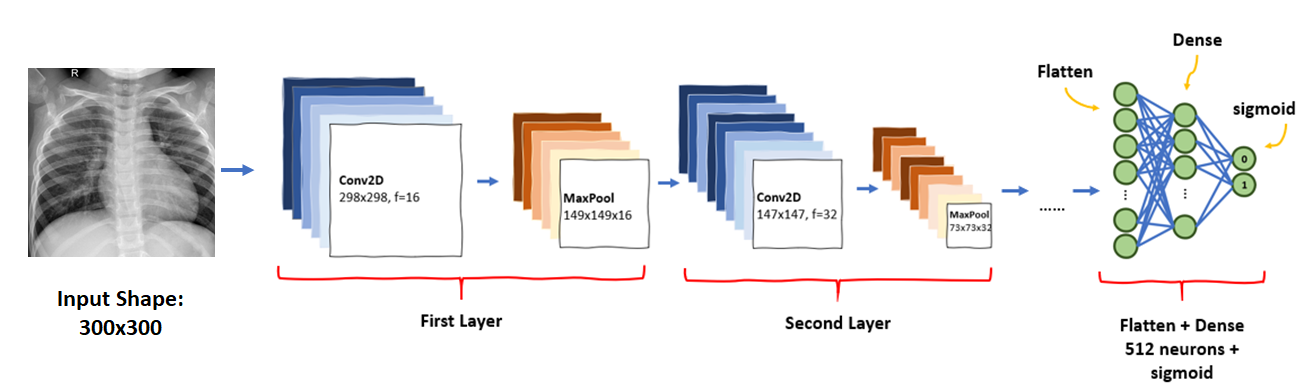}
	\end{tabular}
	\end{center}
  \caption[example] 
  { \label{fig:cnn} 
Sample Convolutional Neural Network architecture followed by non-linear layers and sigmoid function to convert embeddings into probabilistic output of possible categories. } 
  \end{figure} 

Creating an efficient performance model requires two main properties: Good Embeddings and better objective function. For image-based features, a model needs to have a high capacity or learned weights to extract high-level features. Similarly, the objective function should create clearer defined segregation among classes even in case of scarce or biased data.

\subsection{Baseline Model: Logistic Regression}
For this research, we have taken logistic regression as a baseline model for Covid classification. Logistic regression consists of one layer of non-linearity using softmax followed by cross-entropy objective function. The logistic Regression model generally works best when the objective focuses on some low-level feature such as binary classification of a house-sale or color-basis classification.

\subsection{Convolutional Neural Networks based model \cite{lecun1995convolutional}}
Deep Learning has transformed many industries ranging from the manufacturing industry, food industry, and now medical industry. Among all architectures, Convolutional Neural networks \cite{lecun1995convolutional} played an important role in the computer vision domain. Convolutional Neural Networks are inspired by mammals' visual cortex and how their vision system uses a layered architecture of neurons in the brain. Just like humans have a group of neurons to recognize shapes and sizes, Convolutional neural network layers extract specific forms of features from image input to analyze the object in an image. Convolutional layers are also called feature extractor layer because a range of features of the image is extracted within each layer. In this work, we have implemented a ten layered convolutional neural network for Covid classification. For our experiments, we used a 5 layer convolutional neural network followed by 5 linear layers with cross entropy loss function as objective.

\subsection{Transfer Learning \cite{alzubaidi2020towards}}
Transfer learning refers to a scenario where an architecture that has been optimized on a similar domain data-set, can be used to learn a low-data regime objective. It uses one trained neural network for generalization and then uses the current data-set to improve those parameters. Transfer learning \cite{alzubaidi2020towards} became popular and has helped resolve scarce data situations in many cases, but we generally do not get similar domain data with medical images. In this research, we have taken one of the leading VGG-16 architecture trained on Image-Net for transfer learning. We have chosen different domain for pre-training as training a VGG-16 level of architecture from scratch requires more than 10k images properly annotated and access to good computation power. For our experiments, we extracted the features of the final convolutional layer of VGG-16 Net and added a logistic regression layer followed by cross-entropy loss for Covid classification.

  \begin{figure} [!ht]
  \begin{center}
  \begin{tabular}{c} 
  \includegraphics[height=6cm]{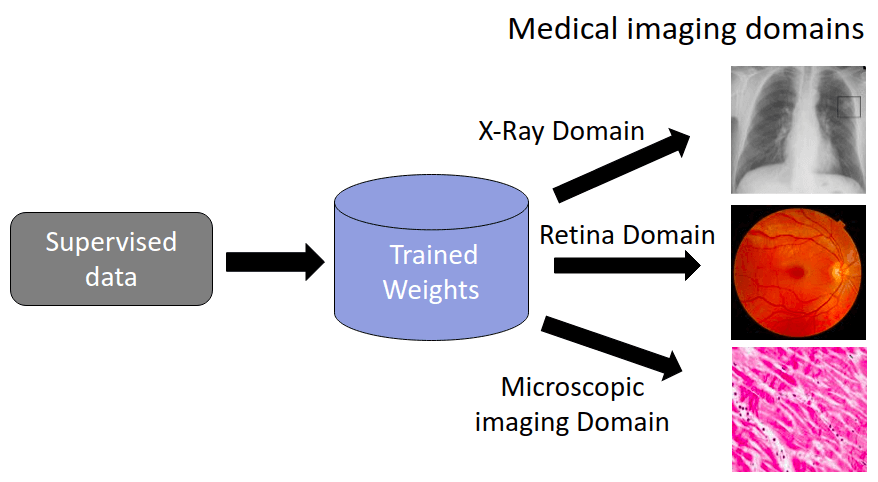}
	\end{tabular}
	\end{center}
  \caption[example] 
  { \label{fig:transfer} 
An Example of how transfer learning can be utlized in various sub-fields of medical industry with help of similar domain supervised data.} 
  \end{figure} 

\subsection{Unsupervised Learning based models \cite{chen2020unsupervised}}
\subsubsection{t-SNE \cite{perez2020improving} and PCA}
\begin{figure} [!ht]
  \begin{center}
  \begin{tabular}{c} 
  \includegraphics[height=8.0 cm]{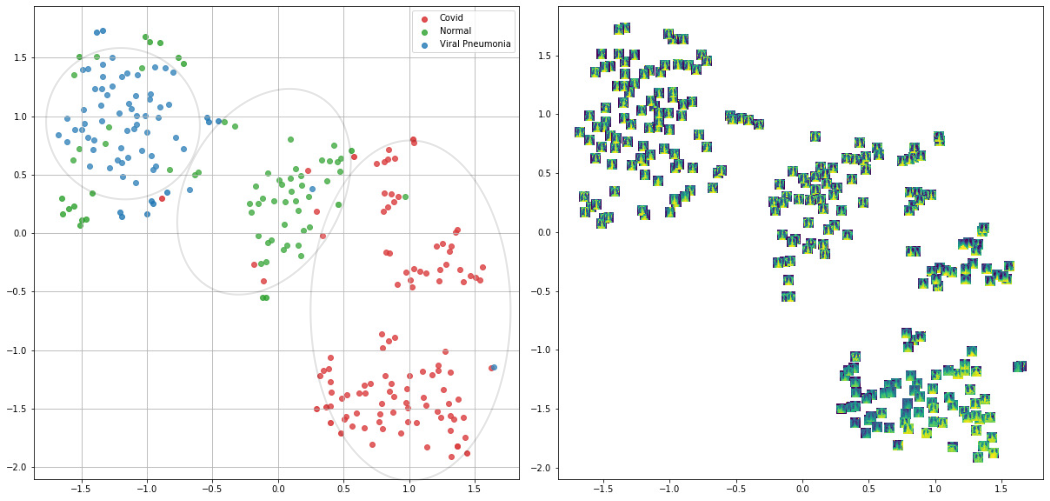}
	\end{tabular}
	\end{center}
  \caption[example] 
  { \label{fig:tsne} 
2-D Visualization of formed clusters using K-Means approach by extracting the feature embeddings using dimensionality reduction approach of PCA followed by t-SNE} 
  \end{figure}

t-Distributed Stochastic Neighbor Embedding (t-SNE) is a widely used technique for dimensionality reduction. It is particularly used for the visualization and analysis of high-dimensional data-sets. t-Distributed stochastic neighbor embedding (t-SNE) minimizes the divergence between two distributions: a distribution that measures pairwise similarities of the input objects and a distribution that measures pairwise similarities corresponding to low-dimensional points in the embedding. In simpler terms, t-SNE analyzes the original data entered into the algorithm and hypothesizes the best representation of data in lower dimensions by matching both distributions. The hypothesizing process is computationally expensive; therefore, there are certain limitations to its usage with real data. For example, in very high dimensional data, to avoid the heavy computation, we can add another dimensionality reduction technique before using t-SNE. In this research, as our data comes under high dimensional data (251X224X224X3), we have used another popular dimensionality reduction approach known as PCA(Principal Component Analysis). PCA attempts to reduce the size of feature dimensions with eigen vectors' help while ensuring we are not losing essential information. For our experiments, we have first used PCA to bring the dimensions down to 180 and later used t-SNE for visualization, as shown in fig \ref{fig:tsne}. \\

\subsubsection{K-Means Clustering}
K-Means clustering is a 3-step process.
\begin{enumerate}
    \item initialize K centroids in the embedding space.
    \item Compute the distance of each point from each centroid and assign a point to that cluster whose centroid is closest to it.  Do this for every point until preliminary clusters have been formed.
    \item  Within each cluster, recompute the centroid and repeat step 2 until clusters stop changing. Here, K is the number of clusters that the user wants.
\end{enumerate}
The objective of K-Means is to partition N data points into K clusters in such a manner that the within-cluster sum of squares or variance is minimized.  We used scikit-learn’s available k-means clustering algorithm for our implementation, with randomly initialized centroids.
\subsubsection{Gaussian Mixture Models Clustering}
In Gaussian Mixture Model clustering, clusters are modeled with Gaussian distributions, which means that we use variance and the mean to define each cluster.GMM allows for overlapping clusters, with the mixture model being parameterized by three values - each cluster's weight, the mean of each cluster, and the variance of each cluster.  The probability of belonging to a particular cluster is assigned to each data point using the Expectation-Maximization algorithm.  Given the number of component gaussians or clusters (K), this algorithm consists of the following two steps:1.  Expectation:  
\begin{enumerate}
\item Expectation: In the first step, the probability of each point belonging to a cluster calculated for the current values of weight, mean, and variance of that cluster.
\item Maximization:   In this step,  the expectation calculated in the previous step is maximized by modifying the values of weight, mean, and variance of clusters.
\end{enumerate}

This iterative model runs until convergence,  at which point the maximum likelihood estimate is provided.  Once the model parameters have been estimated, the fitted model can be used for clustering.  A point is assigned to that cluster for which the probability of it belonging to the cluster is maximum.

\subsection{Few-Shot Learning using Siamese Networks \cite{jadon2021improving}}

Few-shot learning is a sub-field of machine learning which aims to develop models that can be trained with less amount of data-set and provide the required performance. For a model to be efficient, it requires good embeddings or better optimization approach which can reach the desired objective within less steps. There are three types of few-shot learning apporoaches: Metrics based, Models based, and Optimization based; Metric based approaches focuses of learning better embeddings whereas Models based and Optimization based approaches focuses on improving the architectural components and optimization algorithms respectively. In this work, we have taken advantage of one of these Metrics based approach known as Siamese Networks. Siamese stands for `twins`, and as the name suggests Siamese Networks consists of two architectures similar in all features and shares weights among it. For image type data, these similar architectures are generally chosen to be of Convolutional Neural Network type followed by contrastive loss function. For training a siamese network, we pass the input in set of pairs, e.g; we take 2 input images and label them if they are similar or not, then these two input images (x1 and x2) are passed through the ConvNet to generate a fixed length feature vector for each (h(x1) and h(x2)). Assuming the neural network model is trained properly, we can make the following hypothesis: If the two input images belong to the same character, then their feature vectors must also be similar, while if the two input images belong to the different characters, then their feature vectors will also be different. This idea of extracting embeddings on basis of similarity and dissimilarity helps model to train with less number of examples and even take advantage of transfer learning approach.For our research, we further modified the Siamese network to get features followed by VGG16 layers (trained on Image Net).
 
   \begin{figure} [!ht]
  \begin{center}
  \begin{tabular}{c} 
  \includegraphics[height=6cm]{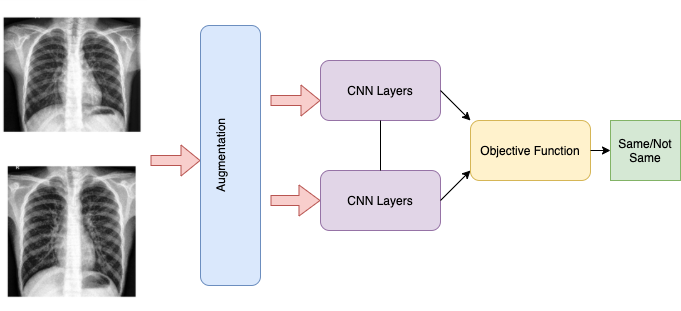}
	\end{tabular}
	\end{center}
  \caption[example] 
  { \label{fig:siamese} 
An architectural representation of Siamese Networks with case of two different class input.} 
  \end{figure}


\section{Evaluation Metrics}
To assess the performance and understand machine learning models, we needed a good set of evaluation metrics. For this work, we have used five metrics: Accuracy, Precision, Recall, F1-score, and the Silhouette score.
  
\subsection{Accuracy}
Accuracy is the most intuitive performance measure, and it is merely a ratio of correctly predicted observations to the total observations. One may think that if we have high accuracy, then our model is best. Yes, accuracy is an excellent measure only when we have symmetric data-sets where false positives and false negatives are almost the same. Therefore, we have to look at other parameters to evaluate the performance of our model.
\begin{equation}
Accuracy=\frac{TP+TN}{TP+TN+FP+FN}
\end{equation}
\subsection{Precision}
Precision is the ratio of correctly predicted positive observations to the total predicted positive observations. For each category/class, there is one precision value. We focus on precision when we need the predictions to be correct, i.e., ideally, we want to make sure the model is right when we predict a label.
\\
\begin{equation}
    Precision=\frac{TP}{TP+FP}
\end{equation}
\subsection{Recall}
Recall is the ratio of what the model predicted correctly to what the actual labels are. Similarly, for each category/class, there is one recall value. We care about recall when we want to maximize the prediction of a particular class, i.e., ideally, we want the model to capture all the class examples.\\
\begin{equation}
Recall=\frac{TP}{TP+FN}
\end{equation}
\subsection{F1 Score}
F1 score is the weighted average of Precision and Recall. Therefore, this score takes both false positives and false negatives into account. Intuitively it is not as easy to understand as accuracy, but F1 is usually more useful than accuracy, especially if we have an uneven class distribution. Accuracy works best if false positives and false negatives have similar costs. If the cost of false positives and false negatives is very different, it’s better to look at Precision and Recall.
\\
\begin{equation}
F1-Score = \frac{2*Precision*Recall}{Precision + Recall}
\end{equation}
\subsection{Silhouette Score}
Silhouette score is used to analyze the clustering-based approaches. It analyzes the mean of intra-cluster vs. mean of inter-cluster distance using the formula: \\
\begin{equation}
    Silhouette Score = \frac{(b - a)}{max(a, b)}
\end{equation}
There is one condition that the number of clusters' should be less than the number of samples - 1. It ranges from -1 to 1, where 1 represents the best value.

\section{Experiments and Results}
In this paper, we have performed experiments using listed models and evaluated them based on well-known metrics listed in the above section. We used the TensorFlow library for the implementation of our models on \\ 
\textbf{CPU:} Intel(R) Core(TM) i7-6700K CPU @ 4.00GHz\\
\textbf{RAM:} 4x 8GB, 2133 MT/s\\
\textbf{GPU:} GeForce GTX 1060, 6GB \\
To validate the effectiveness of the proposed Siamese based approach with transfer learning features, which is inspired by VGG16 architecture. We first explored some image augmentation and increase the data-size by 10\% making our data-set of ~3800 and ~4200, respectively. After further pre-processing such as normalizing and reshaping the CT-scan images to (224X224), we divided the data-set for training 60\%, validation 20\% and testing 20\% which belonged to 3 classes: Covid, Viral Pneumonia, and Normal.

\begin{table}[!ht]
\caption{Accuracy, Precision, Recall and F1-Score using mentioned classification based modeling approaches trained and tested on combined datset-1 and dataset-2.} 
\label{tab:Experiments}
\begin{center}       
\begin{tabular}{|l|l|l|l|l|} 
\hline
\rule[-1ex]{0pt}{3.5ex}  Model Name & Accuracy & Precision & Recall & F1-score \\
\hline
\rule[-1ex]{0pt}{3.5ex}  Logistic Regression & 82.4\% & 0.822  & 0.828 & 0.828\\
\hline
\rule[-1ex]{0pt}{3.5ex}  Convolutional Neural Network & 90.2\% & 0.912 & 0.901 & 0.904 \\
\hline
\rule[-1ex]{0pt}{3.5ex}  Transfer Learning(VGG16) & 93.3\% & 0.931 & 0.932 & 0.928  \\
\hline
\rule[-1ex]{0pt}{3.5ex} Siamese Networks & 94.6\% & 0.945 & 0.941 & 0.947 \\
\hline
\rule[-1ex]{0pt}{3.5ex} Siamese Networks(Transfer Learning) & \textbf{96.4\%} & \textbf{0.965} & \textbf{0.962} & \textbf{0.959} \\
\hline
\end{tabular}
\end{center}
\end{table}


\begin{table}[!ht]
\caption{Analysis of Clusters formed by K-Means and GMM algorithms with Silhouette Score using input as extracted embeddings of mentioned classification approaches.} 
\label{tab:Paper Margins}
\begin{center}       
\begin{tabular}{|l|l|l|} 
\hline
\rule[-1ex]{0pt}{3.5ex}  Model Name / Clustering Approach & K-Means & Gaussian Mixture Models  \\
\hline
\rule[-1ex]{0pt}{3.5ex}  Logistic Regression & 0.156 & 0.158 \\
\hline
\rule[-1ex]{0pt}{3.5ex}  Convolutional Neural Network & 0.165 & 0.171 \\
\hline
\rule[-1ex]{0pt}{3.5ex}  Transfer Learning & 0.189 & 0.185\\
\hline
\rule[-1ex]{0pt}{3.5ex}  PCA+TSNE & \textit{0.578} & \textit{0.575} \\
\hline
\rule[-1ex]{0pt}{3.5ex}  Siamese Networks &  0.490 & 0.487 \\
\hline 
\rule[-1ex]{0pt}{3.5ex}  Siamese Networks (Transfer Learning) &  \textbf{0.592} & \textbf{0.583} \\
\hline 
\end{tabular}
\end{center}
\end{table}
\subsection{Observations}
After performing experiments and analyzing outcomes we have come to two major conclusions on training models in scarce data situations.
\begin{enumerate}
    \item \textbf{Change is data distribution may result in low accuracy}: When data is scarce, it's tough to determine the real distribution of data, even if we segregate data into train, test, and val. We might not be able to capture the performance of model on real data distribution.
    
    \item \textbf{General Classification models might not work}: In case of less data, generally unsupervised based approaches perform well. In our experiments, we have observed that the decision boundaries were more segregated in the PCA+TSNE and Siamese Network approach, whereas for Logistic Regression, CNN, and Transfer Learning score observed is below 0.2 for both K Means and GMM clustering approaches.
\end{enumerate}

\section{Conclusion}
In this research article, we have experimented with deep learning architectures for the low-data regime. We also proposed a custom metrics based few-shot learning model to predict the multi-class classification of Covid-19.  We approve our model with detailed experiments on the combined Covid-19 radiography collected dataset, where CT scan image belongs to 3 categories as Normal, Viral Pneumonia, and Covid is used to obtain the highest accuracy of our proposed approach. We also investigate the reduction of overfitting and regularization of the model effect on our application performance. For this purpose, we used embedding analysis using silhouette score and clustering approaches. Finally, we compare our proposed technique to the existing three widely used classification approaches, where our proposed model significantly performed better than the others. We can see our proposed approach providing a 3\% increment in accuracy for classification and ~0.42 increment in clustering score. In the future, we plan to examine whether the same model can be employed on the other computer-aided diagnostic problems.

\nocite{*}
\bibliography{main} 
\bibliographystyle{spiebib} 

\end{document}